\documentstyle[prb,aps,epsf]{revtex}
\tighten
\begin{document}

\title{Time Dependent Current Oscillations Through a Quantum Dot}

\author{E. S. Rodrigues$^\dagger$, E. V. Anda$^\ddagger$ and P.
Orellana$^{\dagger \dagger}$}
 
\address{$\dagger$ Instituto de F\'{\i}sica, UFF, 
Av. Litor\^anea s/n$^{\underline o}$, Gragoat\'a, Niter\'oi RJ,
Brasil, 24210-340.\\
$\ddagger$ Departamento de F\'{\i}sica, PUC-RJ, 
Caixa Postal 38071, Rio de Janeiro, Brasil, 22452-970.\\
$\dagger \dagger$ Departamento de F\'{\i}sica, Universidad Cat\'olica
del Norte,
Angamos 0610, Casilla 1280, Antofagasta, Chile.}

\maketitle

\section{Abstract}

Time dependent phenomena associated to charge
transport along a quantum dot in the charge quantization  regime 
is studied. Superimposed to the Coulomb blockade behaviour the 
current has novel non-linear properties. Together with static 
multistabilities in the negative resistance region of the I-V 
characteristic curve, strong correlations at the dot give rise
to self-sustained current and charge oscillations. Their properties
depend upon the parameters of the quantum dot and the external applied 
voltages.\\

\noindent PACS numbers: 73.40.Gk, 85.30.Mn

\section{Introduction}

Electron correlation effects on the transport properties of a quantum
dot ({\bf{QD}}) under the influence of an external potential have
been the subject of many experimental and theoretical studies in
recent years.  This device although possesses thousands of atoms
behaves as if it were an artificial atom due to its discrete energy
spectrum. The enormous advantages of this artificial atom are that its properties can be modified
continuously by adjusting  the external potentials  applied to it [1].
Unlike a real atom it can be experimentally isolated
and its transport properties obtained studying  the flow of charges
through it. The quantization of charge and energy
plays the major role in the transport properties of a {\bf QD}.
The conductance exhibits oscillations as a function of the external
potential which can be explained in terms of a transport
mechanism governed by single-electron tunneling and Coulomb
blockade effect due to the e-e interaction inside the dot[2]. 

It is well established that the Coulomb interaction
produces non-linear effects in {\bf 3D} double barrier heterostructure,
observed as intrinsic static bistability in the negative differential
resistance region of the I-V characteristic curve of the device[3]. This
property has been explained as an electrostatic effect due to the e-e
interaction among the charges inside the well. Besides this static
behavior, theoretical studies have proposed that the Coulomb
interaction can produce time dependent oscillations[4]. More recently 
current oscillations promoted by an external magnetic field applied in 
the direction of the current were predicted[5]. These last results showed 
that the system bifurcates as the field is increased and may transit to 
chaos at large enough fields. Recently very interesting experimental 
evidences of the existence of these oscillations induced by an external 
magnetic field, have been given[6]. 

Although Coulomb interaction phenomena have been a matter of great concern in
the last years, to the best of our  knowledge, there have not been experimental 
studies reporting neither static bistabilities nor current and charge oscillations 
through a {\bf QD}.
In the search for stationary behavior, we were
able to theoretically predict a  bistability similar to the one that appears in the I-V
characteristic curve of a {\bf 3D} double barrier heterostructure. We obtained as well evidences 
of the non-existance of stationary solutions in some regions of the parameter space for the
currents flowing through the {\bf{QD}}[7]. 

In this communication letter we address the study of currents
going through a {\bf{QD}} connected to leads under the effect of an 
external potential. We report the existence of static hysteresis behavior
of the current in the negative resistance region. What is more  important, we
find novel time dependent phenomena which appear as self-sustained
current and charge oscillations with properties depending on the
internal parameters of the system and the external applied potentials. 
These time dependent phenomena are derived from the highly 
correlated electrons inside the {\bf{QD}}. 

It has been shown that the physics
associated to a {\bf{QD}} connected to two leads can be readily
understood in terms of a single magnetic impurity Anderson
Hamiltonian, where the impurity is the quantum dot[2]. 
The leads can be represented by a {\bf 1D} tight-binding Hamiltonian 
connecting the dot to particle reservoirs characterized by Fermi levels 
$E_l$ and $E_r$. The difference $E_l-E_r$ gives the bias voltage
applied to the system when connected to a battery. The Hamiltonian is 
given by

\begin{eqnarray}
H &=& \sum_{i \sigma} \epsilon_i n_{i \sigma} + v \sum_{<ij> \sigma}
c^{\dagger}_{i \sigma} c_{j \sigma}
+ \,(\epsilon_o + V_g) \sum_{\sigma}n_{o \sigma}
+ U n_{o \uparrow} n_{o \downarrow}.
\label{eq1}
\end{eqnarray} 

\noindent The diagonal matrix element $\epsilon_i$ is site dependent 
because it takes into account the potential profile and 
the two barriers. The gate voltage $V_g$ is 
supposed to be effective only on the {\bf QD}.
Since we are not interested in very low temperature
physics, as for instance the Kondo effect, we use a decoupling 
procedure[8] to solve the equation of motion for the one particle 
operator. Although this is a simple approximation, it is adequate 
to treat high correlations at the dot, responsible for Coulomb
blockade phenomena[2]. The non-linear behavior stems from the  
dependence of the solution upon the self-consistent 
calculated charge content of the dot. As we are not interested in 
studying magnetic properties, we omit the subindex $\sigma$ corresponding 
to the spin. Using the set of Wannier functions localized 
at site i, $\phi_{i} (r)$, we can write 
the time dependent state $\varphi_{k} (r,t)$ as,

\begin{equation}
\varphi_{k} (r,t) = \sum_i a^k_{i} (t)\phi_{i} (r)
\label{eq2}
\end{equation}

The probability amplitude for an electron to be at site $j$ in a time dependent 
state $\varphi_{k} (r,t)$ is given by the equation of motion,

\begin{equation}
i \hbar \frac{da^k_{j}}{dt} = \epsilon_j a^k_{j} + v(a^k_{j-1} +
a^k_{j+1})\;\;\;\;\;\;\; (i \not= 0)
\label{eq3}
\end{equation}

\begin{equation}
i \hbar \frac{da^{k,\alpha}_{o}}{dt} = (\epsilon^{\alpha}_o +
V_g) a^{k,\alpha}_{o} + n^{\alpha} v(a^k_{-1} + a^k_{1}) 
\label{eq4}
\end{equation}

\noindent where $\alpha = \pm$; $n^+ = n$; $n^- = 1 - n$; 
$ \epsilon^+_o = U$; $\epsilon^-_o= 0$ and $n$ is the number 
of electrons per spin at the dot that is 
calculated from the probability amplitudes by the equation,

\begin{equation}
n = 2 \int_0^{k_f}  \mid a_{o}^k \mid^2 dk
\label{eq5}
\end{equation}

\noindent where $a_{o}^k = a_{o}^{k,+} + a_{o}^{k,-}$ 
and $k_f$ is the Fermi wave vector of the emitter side.
We have assumed for
simplicity that the collector is not doped so that the charge that
goes towards the emitter is restricted to the electrons that have
been reflected on the barriers.  The time dependent equations are
solved by discretizing equations (3) and (4), and using a half-implicit 
numerical method, which is second-order accurate and unitary[9].
 
The solution of these equations require the knowledge of the wave function
at $-\infty$ and $+\infty$. We suppose that an incident 
electron of wave vector k is described by the function,

\begin{equation}
a^k_{j} = (I e^{ikx_j} + R_j e^{-ikx_j})e^{-i\epsilon t/\hbar}\;\;\;\;\;\; 
(x_j \le -L) 
\label{eq6}
\end{equation}

\begin{equation}
a^k_{j} = T_j e^{ik^{\prime}x_j}e^{-i\epsilon t/\hbar}\;\;\;\;\;\; 
(x_j \geq L) 
\label{eq7}
\end{equation}

\noindent where the system is explicitly defined between $-L$ and $L$
and $k^{\prime}$ is the wave vector of the transmitted particle 
$k^{\prime} = \sqrt(\frac{2m(E_l-E_r)}{\hbar^2}+k^2)$. 
The incident amplitude I is assumed to be spatially constant.  Instead, 
far from the barriers, the envelope function of the reflected and 
transmitted waves $R_j$ and $T_j$ 
are supposed to be weakly dependent on site $j$. This permits
to restrict the dependence to the linear term, which results to be an
adequate approximation provided the time step taken to discretize
equations (3) and (4) is less than certain limit value that depends
upon the parametrization of the system. A maximum value of $0,3 fs$
was adequate to guarantee numerical stability
up to 200 ps. To eliminate spurious reflections at the boundary
it was necessary to take systems of the order of $2L=400$ sites. 
In the numerical procedure, the Wannier intensities obtained for 
one bias are used as the starting point for the next bias step. Once 
the Wannier intensities are known the current is calculated from[9],

\begin{equation}
J_j \propto 2 \int^{k_f}_o \biggl[
Im(a^{k^*}_{j} (a^k_{j+1} - a^k_{j}) \biggr] dk
\label{eq8}
\end{equation}

It is important to emphasize that the system we study has to be 
in the Coulomb blockade regime. Charge quantization is obtained 
satisfying the relation $\frac{t^{\prime 2}}{W} < U$ where $t^{\prime}$ 
is an effective coupling constant between the 
{\bf{QD}} and the leads controlled by the barriers parameters
and W is the leads bandwidth. In this case the energy spacing U
between the states with $N$ and $N+1$ electrons is greater than the width
of the resonant levels within the dot.

We study the system defined 
by an emitter barrier of $1.4 eV$ and a collector barrier of
$2.4 eV$ both of a thickness of 5 lattice parameters and an electronic
repulsion  U = 20meV. This configuration represents a GaAs structure 
in the charge quantization regime. 
A similar effective coupling constant can be obtained assuming lower
and thicker barriers. We have chosen the case with less number of
sites in order to reduce the computational effort. 
In order to enhance the charge content of the dot and consequently the
non-linear effects the second barrier is taken to be higher than the
first. The Fermi level lies 30 meV above the bottom of the emitter
conduction band.
With these parameters and zero gate voltage and bias the localized 
level at the dot lies well above the Fermi level, so that the system 
is completely out of resonance. For small values of the gate potential 
equations (3) and (4) have a stationary solution after a transient, 
for the whole interval of bias voltage $V$, which drives the system 
into and off resonance. Increasing the gate potential the
system enters into a completely different regime. As the bias 
voltage required to drive the system into resonance is small,
because the resonant level is now near the Fermi level, the
second barrier is not significantly reduced by the bias voltage
and is able to trap a large amount of electronic charge inside the
dot. As a consequence, the non-linear effect is enhanced giving
 rise to a region in the I-V characteristic curve where there is
no stationary solution. In this parameter region the current has 
self-sustained oscillations while for greater values of the bias
voltage the system possesses  complex electrostatic multistabilities 
which can be obtained, depending upon the initial condition, augmenting 
or diminishing the bias voltage. These static multistabilities are very 
well known phenomena of {\bf 3D} systems, although they generally assume 
a bistable structure[3]. The situation for the {\bf{QD}} described is 
illustrated in Fig. 1 where the bubble shows the maximum and minimum of 
the current oscillations, while for a greater value of the bias
voltage the electrostatic instabilities are clearly represented.
The non-stationary solutions lie outside the static multistable region for
small values of $ V_g$. However, for larger values of $ V_g $, there is
a superposition of the regions where both effects are active.
The static large multistability is related to the drop of the resonant 
level below the bottom of the conduction band, while the dynamic 
oscillations appear when the resonant level aligns with the Fermi level. 
The size and location of the oscillating behavior in the I-V space depends 
upon the gate voltage. Its size increases with $ V_g $.
The non-linearities  are enhanced by the gate potential,
constituting an external tunable parameter through which the
oscillating phenomena can be controlled. 
\epsfxsize=220pt
\begin{figure}[htbp]
\centerline{\epsfbox{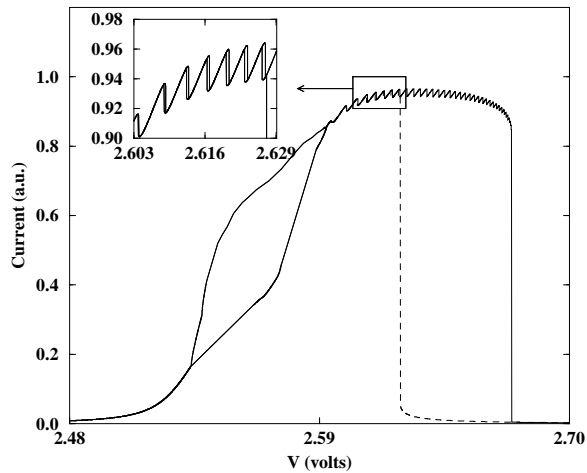}}
\caption{Current versus bias voltage, I-V characteristic curve. The bubble 
indicates the maximum and minimum of the oscillating current amplitude. The 
inset shows a complex structure of static bistable behaviour. Dashed line shows
the uncharged {\bf QD} solution.}
\label{fig1}
\end{figure}

We have taken different values 
of the bias  $ V_g $ to study explicitly the time dependent current going
along the device. For $V_g = 0.2$ volts the system possesses a stationary 
solution for all values of the bias voltage. 
For a greater value of the gate 
potential $V_g = 0.86$ volts and $V = 5.6$ volts, we show the current 
and the charge as a function of time in three different locations of 
the system: in the emitter and collector side before and after the 
barriers, in Fig. 2b and,  inside the dot, in Fig. 2c. We see that 
after a transient of some ps, the system begins to oscillate in a
self-sustained way.  It is important to note the different behaviors
of the oscillations depending upon the site where the current is
evaluated.  The oscillations are large in the region before the
emitter barrier and inside the well, while after the charge has
passed through the device the more opaque collector barrier damps
the oscillation in a very significant way. Although not perfectly
periodic, the oscillations have some regularity. The frequency
spectrum is continuous with the predominance of two and some times
three well define frequency regions. The origin of this behavior is not
completely clear to us. However, the existence of more than one frequency 
region has to be attributed to the amount of non-linearity in the system. A
less non-linear situation, obtained for instance in the near vicinity of
the bifurcation point where time dependent behavior appears in the
I-V characteristic curve of Fig. 1, shows small amplitude oscillating
currents of almost only one frequency.  Very remarkably, superimposed
to the oscillating currents, we exhibit in the inset of Fig. 2c  very
high frequency oscillations greatly enhanced inside the well and
almost non-existent outside it. We attribute them to the electrons
going back and forth inside the well with a period corresponding to
the time taken by the electrons at the neighborhood of the Fermi energy being
successively reflected by the barriers.
\epsfxsize=220pt
\begin{figure}[htbp]
\centerline{\epsfbox{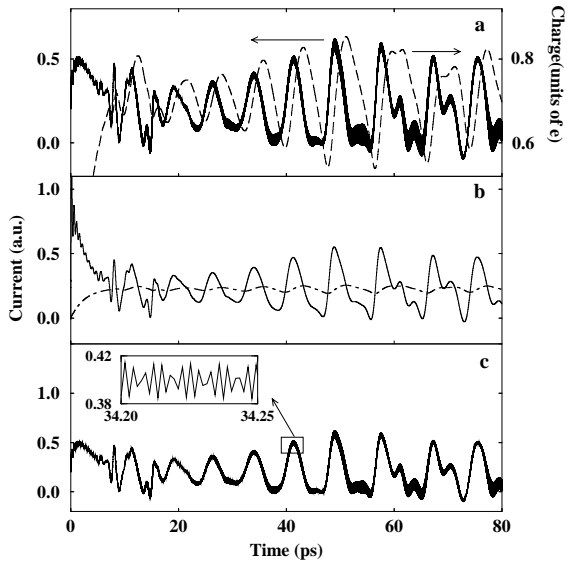}}
\caption{Current and charge versus time for $V_g = 0.86$ volts and $V = 5.6$ volts.
a) Charge (dashed line) and current (full line) at the dot as a function of
time, showing the time lag between them. b) Current at the emitter (full line) and at 
the collector (dashed line). c) Current at the dot. Inset shows high frequency
oscillations.} 
\label{fig2} 
\end{figure}
Changing the gate potential the dot enters into resonance when its local level 
aligns with the Fermi level. The current increases abruptly and, due to the 
trapping effect of the collector barrier, the dot charge  per spin $n$ 
assumes its maximum value after a time. As a consequence, the strength of the 
resonant level, which due to many body effects depends upon the dot charge as  
$(1-n)$, diminishes and the flow of electrons going into the dot reduces. 
As time evolves, the total amount of charge inside the dot drops because the 
leaking of charge through the second barrier becomes greater than its entrance 
through the first one. The reduction of charge increases the strength of the 
resonant level, which increases the flowing current starting a new cycle. 
As we have numerically verified, for a configuration outside the Coulomb
blockade regime, by diminishing for instance the first barrier, the oscillations 
are damped out and the system reaches a stationary regime. In this case, as the system is driven 
into resonance, the current increases and the dot charge reacts almost 
instantaneously with no significant lag time between them. 

To generate self-sustained current oscillations the system has to be in a
parameter region where there is no dynamic accommodation
between the current and the charge inside the dot. It requires an abrupt 
entrance in resonance when the external potential is modified. This is the case 
of the charge quantization regime when the channel for the electron to go 
through is provided by the very sharp resonant peak inside the dot.   
It is interesting to compare the current and the charge oscillations
inside the well in this case.  Although the frequencies and the general 
properties of both oscillating quantities are the same, there is a clear lag
time $\tau$ of the dot charge in relation to the current. The dephasing between 
charge and current which disappears with more transparent barriers is illustrated 
in Fig. 2a. It is clear that the lag $\tau$ depends upon the time scale controlling 
the entrance of the charge into the resonant state, which is called the buildup time.
For opaque emitter barriers, as in our case, the buildup time is
found to be essentially independent of the barrier parameters. The
buildup process is essentially determined by the spatial distribution of the extra charge 
in conditions to enter into the well[10]. This is slightly different in each 
oscillation, creating a rather irregular lag among the two quantities as time 
goes along, as exhibited in Fig. 2a. 

Due to  computer time limitations it was 
not possible to develope a systematic investigation of the phenomenon. However, 
we have studied the system modifying some of its parameters in order to give 
support to the interpretation above. There is a clear tendency of the oscillation 
mean value frequency to increase reducing the opacity of the barriers although 
it is more clearly dependent upon the second barrier. This is 
consistent with the interpretation given above as it is the collector barrier 
that controls the leaking time of the charge when the its entrance is abruptly 
reduced by non-linearity. In order to study the stability of the self-sustained
oscillations we have calculated, for one particular case, the time
evolution of the system up to a value of $200$ps. The oscillations
maintain their amplitude and  regular shape. We have taken  a 
charged and a discharged dot as two different  initial conditions. The first 
condition creates oscillations  immediately after the circuit is switched on 
while in the second the system  goes through a transient
during which the dot is charged, as shown in Fig. 3. Although the oscillations 
have a similar behavior shifting one relative to the other  by this charging
time, there are differences which in fact increase in time as shown
in the inset of Fig. 3. In certain regions of the parameter space the 
oscillations result to be very sensitive to initial conditions, suggesting the 
existence of chaotic phenomena.
\epsfxsize=220pt
\begin{figure}[htbp]
\centerline{\epsfbox{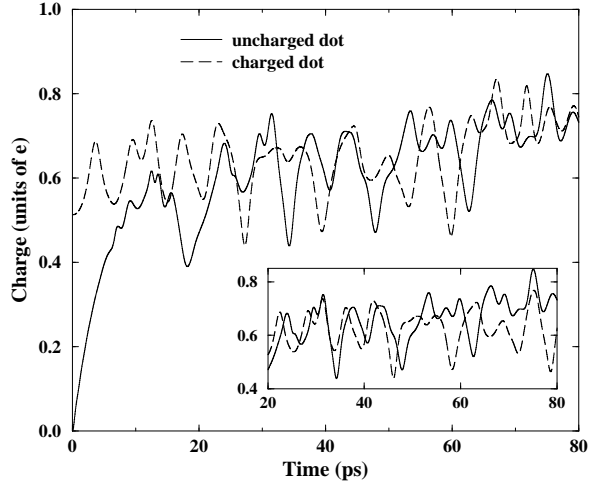}}
\caption{Charge at the dot as a function of time, $V = 3.13$ volts and $V_g =
3.02$ volts. Initial states, dot uncharged (full line) and dot charged (dashed
line). Inset shows the two solutions where the dot uncharged initial 
state solution is shifted by the changing time.}
\label{fig3}
\end{figure}
We have done a time dependent study of the current circulating through a
{\bf QD} in the Coulomb blockade regime. We conclude that together with
complex static mutistable behavior, non linearity produced by highly
correlated electrons in the dot give rise to charge and current time
dependent oscillations. They can be controlled by changing the
parameters defining the system and the various external potentials
applied. A complete characterization of the time dependent oscillations 
would require a great computational effort, which could be justified only 
in the case of an experimental verification of the phenomena  theoretically 
predicted in this letter.

Work supported by FONDECYT
grants 1980225 and 1990443, CONICYT,Catedra Presidencial en Ciencias (Chile),Fundaci\'on Antorchas/Vitae/Andes grant A-13562/1-3
and the brazilians agencies FINEP and CNPq.

\end{document}